\documentclass[12pt]{article}
\usepackage{amssymb}
\usepackage{amsmath}
\usepackage{graphics,epsfig,color,soul}
\usepackage{bm}
\usepackage{color}
\usepackage{authblk}
\usepackage{wrapfig}
\usepackage[export]{adjustbox} 
\usepackage{subcaption}
\usepackage{comment}

\DeclareRobustCommand{\divby}{%
  \mathrel{\vbox{\baselineskip.65ex\lineskiplimit0pt\hbox{.}\hbox{.}\hbox{.}}}%
}

\usepackage{cite}
\usepackage{hyperref}
\hypersetup{
  colorlinks   = true, 
  urlcolor     = blue, 
  linkcolor    = blue, 
  citecolor   = black 
}

\usepackage[export]{adjustbox} 

\usepackage{framed} 
\begin{document}

\title{Biomechanics of orientationally ordered epithelial tissue
}
\author{Patrick W. Alford$^{1}$, Luiza Angheluta$^{2}$, and Jorge Vi\~nals$^{3}$ \\
$^{(1)}$ Department of Biomedical Engineering, University of Minnesota, Minneapolis, MN 55455 \\
$^{(2)}$ Njord Centre, Department of Physics, University of Oslo, P.O. Box 1048, 0316 Oslo, Norway, and \\ $^{(3)}$ School of Physics and Astronomy, University of Minnesota, Minneapolis, MN 55455.}

\maketitle

\begin{abstract}
Organogenesis involves large deformations and complex shape changes that require elaborate mechanical regulation. Models of tissue biomechanics have been introduced to account for the coupling between mechanical response and biochemical processes. Recent experimental evidence indicates that the mechanical response of epithelial tissue is strongly anisotropic, with the degree of anisotropy being correlated with the existence of long range orientational order of cytoskeletal organization across the tissue. A theoretical framework is introduced that captures the dynamic feedback between tissue elastic response and cytoskeletal reorganization under stress. Within the linear regime for small and uniform applied strains, the shear modulus is effectively reduced by the nematic order in cytoskeletal alignment induced by the applied strain. This prediction agrees with experimental observations of epithelial response in lithographically patterned micro tissues. 
\end{abstract}

\section{Introduction}
\label{sec:introduction}

Because of large deformations and complex shape changes, organ formation during development is fundamentally a mechanically driven process. In particular, incorrect structural organogenesis is a key factor in many congenital birth defects. We aim to contribute to our understanding of how complex loads on developing tissues might act to redirect morphogenesis. This requires the development of quantitative physical and mathematical models of mechano-adaption describing how biomechanical forces are involved in both growth and shape change. 

It is well established that active mechanical coupling between the cell and its environment is key to tissue homeostasis, as well as to cell differentiation and tissue remodeling in response to mechanical cues \cite{re:engler06,re:trepat09,re:shyer13,re:alford11}. Our focus here is on the class of tissues that contain regular fibrous matrices which, when spatially oriented, correlate with anisotropic elastic response \cite{re:sacks03,re:humphrey90,re:kim10,re:win17}. Existing phenomenology does suggests that such an anisotropic elastic response, which is coupled to internal cell organization, plays an important role in cell mechanobiology \cite{re:win17,re:cook23}. We introduce a theory that includes elastic anisotropy within the cell, the degree of order in the orientation of its fibrous structure, and the dynamical coupling of both under load
\cite{re:win17,re:steucke17,re:win18,re:rothermel20,re:cook23}. The theory is motivated by the simultaneous measurements of elastic response and fiber orientation in the recently developed cellular micro-biaxial stretching system (C$\mu$BS) \cite{re:win17,re:cook23}, coupled to traction force microscopy. The setup allows the determination of pointwise stresses in the tissue as it is being stretched. It can then be readily correlated with fiber orientation maps obtained through confocal microscopy.

The most widely used class of coarse grained models of confluent tissue mechanical response are the so called vertex models (see, e.g., Ref. \cite{re:alt17} for a review). They focus on the network of vertices that define apical cell surfaces in planar ephitelia, and assume that the mechanical response of the tissue is due to cell edge distortion, with the main forces generated along the cell surfaces. These forces depend only on the surface area of each cell relative to a preferred area, with an energy penalty that depends on  the sum of the edge perimeters. A recent generalization of vertex models are the so called multi phase field models \cite{re:palmieri15,re:mueller19,re:zhang20,re:jain23}. A two dimensional confluent cell monolayer comprising $N$ cells is modeled by $N$ scalar phase field functions $\phi_{i}(\mathbf{x},t)$ where $\phi_{i}$ is chosen to be an indicator function, equal to 1 well inside the bulk of cell $i$, and zero outside. The indicator function evolves in time so as to minimize a governing free energy functional that preserves the volume of each cell, maintains confluence in the tissue, and endows each interface between two adjoining cells with an interfacial tension that depends on the local curvature of the interface. Triple junctions emerge spontaneously, and are treated with the same variables and free energy. Each individual cell is treated as an active (self propelled) element with a predefined velocity that depends on a new dynamical but stochastic variable, the cell polarity.  This last element models the propensity of cells to migrate in a given direction which, while stochastic, it depends on its environment. In both types of models, however, the cell interior is coarse grained away, and therefore any internal cytoskeletal structure does not contribute to the governing free energies, nor to the derived equations of cell motion. 

A complementary approach to tissue biomechanics involves treating both the cell (and an extracellular matrix if present) as continuum, elastic materials \cite{re:fung67,re:humphrey90,re:humphrey03}. Recent models add some degree of cellular microstructural detail into improved constitutive laws by considering, for example, fiber orientation distributions to capture anisotropic behavior of the cell, or the extra cellular matrix surrounding the cells \cite{re:gasser06,re:lai13}. Of particular relevance to our study, is the introduction of the so called \lq\lq anisotropy tensors\rq\rq~to account for structural cytoskeletal orientation within a tissue \cite{re:gasser06,re:barocas97,re:zahalak00,re:marquez05,re:win17}. Finite Element computational frameworks are then utilized to analyze elastic response under a variety of conditions \cite{re:breuls02,re:win17}.

It has been recently established, however, that the degree of cytoskeletal fiber orientation in tissue is itself a dynamical variable, exhibiting spatio temporal variation as the tissue remodels under imposed loads \cite{re:win17,re:cook23}. This observation, together with a measured change in tissue stiffness under stretching along the predominant direction of the fiber network suggest modeling this class of tissue as a uniaxial solid (also known as a transverse isotropic solid), with both elastic distortion and fiber orientation coupled as dynamical fields. This phenomenology is not unlike that of liquid crystal elastomers \cite{re:warner07}. There small molecule monomers or side chains in a polymer backbone define a preferred local orientation at low temperatures (a nematic phase), which affects the elastic response of the polymer matrix. Imposed tractions can lead to molecular reorientation, a dissipative process that absorbs some of the energy imparted. Therefore, molecular reorientation results in the so called \lq\lq soft elasticity\rq\rq, a characteristic of nematic elastomers.

The mechanical response of orientationally ordered tissue likely follows from a combination of the two phenomena described above. The existence of a preferred fiber direction orientation is expected to result in uniaxial elasticity. However, the degree of orientational order is itself coupled to any existing stresses, not unlike the case of nematic elastomers. In addition, fiber network remodeling and cell proliferation in response to stresses are also expected to be important in determining the elastic response. The model introduced here is based on the ideas of continuum mechanics, but allows for a dependence of the governing energy on a spatio and temporally dependent degree of orientation, the nematic order parameter. The coupling between the two phenomena originates in tissue growth, either largely isotropic because of cell proliferation, or strongly anisotropic and related to the degree of nematic order due to local remodeling of the fiber network. Our model accommodates both uniaxial elasticity and nematic elastomer response, and it makes predictions about pointwise distributions of stresses and orientational order. These predictions can be verified by existing traction force microscopy experiments, when coupled to the determination of the tissue fiber orientation distribution.

\section{Transport model}
\label{sec:transport_model}
We consider a deformed tissue due a displacement vector $\bm{u} =\bm x-\bm X$ from the reference state $\bm X = \{X_i\}$ to the deformed (actual) state $ \bm x =\{x_i\}$. The distortion is defined by the gradient $U_{ij} = \partial_{j} u_{i} = \delta_{ij} - F^{-1}_{ij}$, where $F_{ij} = \frac{dx_i}{dX_j}$ is the Jacobian matrix of the coordinate transformation which is invertible for affine (compatible) transformations, i.e. $\det(\bm I-\bm U) \ne 0$. In such cases, the inverse of Jacobian matrix is a direction measure of the distortion, i.e. $\bm W = \bm F^{-1} = \bm{I} - \bm{U}$. The local state of a tissue under distortion depends on the distortion $\bm U = \bm I-\bm{W}$. The state of fiber alignment is represented by the nematic tensor order parameter field $\bm{Q}$. 

In order to characterize the local state of fiber orientation, one can introduce the so called nematic director field, a unit vector $\hat{\bm{n}} (\mathbf x)$ that gives the local statistical average of the orientation of the fibers \cite{re:degennes93,re:selinger16}. Figure \ref{fi:op} shows a bright field image of a patterned micro tissue stained for its F-actin network, together with the predominant fiber orientation $\hat{\bm{n}} (\mathbf x)$. The figure also shows the distribution of fiber orientations, for three different tissue patterns and state of elongation (see Ref. \cite{re:cook23} for details and protocols). The director only determines direction, and hence the orientation distribution is invariant under $\hat{\bm{n}} \rightarrow -\hat{\bm{n}}$. An alternative representation of orientational order is the nematic tensor order parameter $\bm{Q}(\bm{x})$ which gives {\em both} the magnitude and direction of local fiber orientation. If $p(\bm{\xi})$ is the local distribution of fiber orientation, the tensor $\bm{Q}$ is defined in terms of  the average
$$
\bm{Q} = \int_{\mathcal{S}^{2}} d \Sigma \ (\frac{3}{2}\bm{\xi} \otimes \bm{\xi} - \frac{1}{2} \bm{I}) p(\bm{\xi{)}}
$$
The domain of integration is the unit sphere $\mathcal{S}^{2}$, with surface element $d \Sigma$. By definition the tensor $\bm{Q}$ is symmetric and traceless. $\bm{Q}$ may be diagonalized with real eigenvalues, $\lambda_{1} \ge \lambda_{2} \ge \lambda_{3}$ , and corresponding orthonormal eigenvectors , $\hat{\bm{n}}$, $\hat{\bm{m}}$, $\hat{\bm{l}}$ . The well known scalar uniaxial order parameter (the measure of uniaxial order) may be defined as $S = \lambda_{1}$. 

Under isothermal conditions, the free energy of a deformed body with some degree of nematic order is written as
\begin{equation}\label{eq:free_energy}
{\mathcal F} \left[ \bm{W}(\bm{x},t), \bm{Q}(\bm{x},t), \nabla \bm{Q}(\bm{x},t) \right] = \int_{\Omega} dV \ \rho f \left( \bm{W}(\bm{x},t), \bm{Q}(\bm{x},t), \nabla \bm{Q}(\bm{x},t) \right),    
\end{equation}
where $\rho$ is the mass density, and $f$ the free energy per unit mass. The free energy can be written as \newline $f \left( \bm{W}(\bm{x},t), \bm{Q}(\bm{x},t), \nabla \bm{Q}(\bm{x},t) \right) = f_{el}( \bm{W}(\bm{x},t), \bm{Q}(\bm{x},t)) + f_{n} (\bm{Q}(\bm{x},t), \nabla \bm{Q}(\bm{x},t))$, the sum of an elastic contribution $f_{el}$ dependent on distortion and the local degree of orientational order, and a nematic contribution $f_{n}$ which we will take of the Landau-de Gennes form \cite{re:degennes93,re:selinger16},
\begin{equation}
\rho f_{n} = \frac{1}{2} K |\nabla \bm{Q}|^{2} - \frac{a}{2} {\rm Tr}(\bm{Q}^{2}) - \frac{b}{3} {\rm Tr}(\bm{Q}^{3}) + \frac{c}{4} \left( {\rm Tr}(\bm{Q}^{2}) \right)^{2} 
\label{eq:ldg}
\end{equation}
where $K, a, b, c$ are material parameters and positive. Minimizers of $f_{n}$ in the absence of distortion are either uniaxial or isotropic $\bm{Q}$ tensors.

Consider now a dissipation inequality according to which the work done by external forces on the body $\Omega(t)$ (the actual body, which is distorted) has to be larger that the rate of change of its free energy plus the kinetic energy. We do not consider below the work done on the external boundary by a time dependent $\bm{Q}$ tensor on the boundary (see \cite{re:degennes93} for an analysis of this contribution). We assume that the tensor $\bm{Q}$ is independent of time on the body boundary. The dissipation inequality reads,
\begin{equation}
\int_{\partial \Omega} ( \bm{T} \cdot \bm{n}) \cdot \bm{v} \ dS \ge \frac{d}{dt} \int_{\Omega} \rho f dV + \frac{d}{dt} \int_{\Omega} \frac{1}{2} \rho v^{2} dV
\end{equation}
where $\bm{T}$ is the Cauchy stress tensor, $\bm{n}$ the unit normal at $\partial \Omega$, the boundary of the body, $\rho$ is the mass density, $\bm{v}$ the local center of mass velocity. In components, contractions follow the convention $T_{ij}\partial_{j}v_{i} = \bm{T} : \nabla \bm{v} = \bm{T}:\bm{L}$, where in the tensor $L_{ij} = \partial_{j}v_{i}$, both velocity and derivatives are spatial, that is, in the actual, deformed body. Standard manipulations using the divergence theorem and the momentum balance $\rho \dot{\bm v} = \nabla\cdot \bm T$, lead to \cite{re:tadmor12,re:taber04}
$$
\int_{\partial \Omega} (\bm{T}\cdot \bm{n}) \cdot  \bm{v} \ dS = \frac{d}{dt} \int_{\Omega} \frac{1}{2} \rho v^{2} dV + \int_{\Omega} \bm{T}: \bm{L} \ dV
$$
such that the energy dissipation inequality reduces to 
\begin{equation}
\int_{\Omega} dV 
\left[\bm{T}:\bm{L}- \frac{d}{dt}  (\rho f) - (\rho f) {\rm Tr} (\bm{L})\right] \ge 0,
\end{equation}
where the last term comes from volume change. 

By taking the material derivative of $\bm{F}\bm{F}^{-1} = \bm{I}$, one finds $\dot{\bm{W}} + \bm{W}\bm{L} = 0$ (for a compatible deformation),  where $\dot{( \; )}\equiv \frac{d}{dt} ()$ is the material derivative $\dot{( \; )} = \partial_{t} ( \; ) + \mathbf{v} \cdot \nabla (\; )$. This relation is standard and follows from the kinematics of the Jacobian matrix $\bm{F}$ \cite{re:tadmor12}, given by $\dot{\bm{F}} = \bm{L}\bm{F}$. The dissipation inequality can be now written as,
\begin{equation}
\int_{\Omega} \ dV \ \left\{\bm{T}:\bm{L} - \left[ \frac{\partial (\rho f)}{\partial \bm{W}}:\dot{\bm{W}} + \rho f {\rm Tr}({\bm{L}}) + \frac{\partial (\rho f)}{\partial \bm{Q}} : \dot{\bm{Q}} + \frac{\partial (\rho f)}{\partial (\nabla \bm{Q})} \divby  
\dot{(\nabla \bm{Q})}\right]\right\} \ge 0.
\end{equation}
The triple dot is a triple index contraction from inner to outer indices. Concerning the last term in the right hand side, we note the material derivative and the partial spatial derivative do not commute. If $\varphi$ is a scalar function, or a component of a vector or tensor, one has \cite{re:gurtin96} $ \partial_{i} (\dot{\varphi}) = \dot{(\partial_{i} \varphi)} + (\partial_{i} v_{j})(\partial_{j} \varphi)$,  which for the $\bm{Q}$ tensor derivative reads \cite{re:tovkach17},
$$
\dot{(\partial_{k} Q_{ij})} = \partial_{k} (\dot{Q}_{ij}) - (\partial_{k}v_{l})(\partial_{l}Q_{ij}) 
$$
The dissipation inequality now reads,
\begin{eqnarray*}
\int_{\Omega(t)} dV \left\{ \left[ T_{ij} + \frac{\partial (\rho f)}{\partial W_{mi}} W_{mj} - \rho f \delta_{ij}  +  \frac{\partial (\rho f)}{\partial(\partial_{j}Q_{mn})} (\partial_{i}Q_{mn})\right]\partial_{j} v_{i}
-  h_{ij} \dot{Q}_{ij} \right\} \ge 0,
\end{eqnarray*} 
where the molecular field $h_{ij}$ is the conjugate of $Q$ from the free energy Eq. \eqref{eq:free_energy} is given by 
\begin{equation}
    h_{ij} = \frac{\delta\mathcal F}{\delta Q_{ij}} =\frac{\partial (\rho f)}{\partial Q_{ij}} - \partial_{k} \frac{\partial (\rho f)}{\partial(\partial_{k}Q_{ij})}.  
\end{equation}
We have used integration by parts and that fact that the tensor $\bm{Q}$ is independent of time on the boundary.

One considers first reversible motion in which the equality holds. For reversible motion, $h_{ij}=0$ and the first term in square brackets needs to be zero. This leads naturally to the Ericksen stress, the reversible component of the stress,
\begin{equation}
T_{ij}^{R} \equiv T_{ij}^{E} =  - \frac{\partial (\rho f)}{\partial W_{mi}} W_{mj}  -\frac{\partial (\rho f)}{\partial(\partial_{j} Q_{mn})} \partial_{i}Q_{mn}+ \rho f \delta_{ij}.   
\label{eq:ericksen}
\end{equation}
The second term is the analog of the capillary stress $\frac{\partial f}{\partial (\partial_{i} \psi)} \partial_{j} \psi$ for a scalar order parameter $\psi$ in the Cahn-Hilliard fluid theory\cite{re:gurtin96}.

Before we proceed with dissipative contributions, we note the disparity in time scales between the elastic response of the tissue, tissue shape remodeling through actin polymerization (on the order of seconds), and cell proliferation in response to stresses (on the order of hours). This is clearly illustrated in traction force microscopy experiments in single cell tissue \cite{re:win17}, and in patterned microtissue \cite{re:cook23}. Therefore we adopt a quasistatic approximation according to which the material is in elastic equilibrium compatible with the instantaneous conditions of fiber orientation and cell number, $\nabla \cdot \bm{T}^{R} = 0$. As discussed below, and in order to develop a computational method of the transport model, we note that the governing equations depend both on the distortion, and center of mass velocity. For compatible motion one has $\mathbf{v} = \dot{\mathbf{u}}$. The condition of elastic equilibrium can be re written as $\partial_{j} \dot{T}_{ij} = \partial_{j}(\partial_{t} T_{ij} + v_{k} \partial_{k} T_{ij}) = 0$ since $\partial_{j}T_{ij} = 0$. Therefore the equation of elastic equilibrium can be written in terms of the velocities as well as $\nabla \cdot \dot{\bm{T}} = 0$. We do not include viscous dissipation in the momentum balance equation.

Concerning the orientation order parameter contribution to the dissipation inequality, it can be guaranteed if one chooses,
\begin{equation}
\overset{\circ}{Q}_{ij} = - \Gamma h_{ij},
\label{eq:qevol}
\end{equation}
with $\Gamma >0$ a rotational diffusion constant. In continuum mechanics \cite{re:sonnet12,re:bouck24}, the material derivative appears in the left hand side of Eq. \eqref{eq:qevol}. However, the explicit introduction of additional dissipative terms in velocity gradients requires consideration of the corototational derivative $\overset{\circ}{\bm{Q}}$. The Rayleigh dissipation function needs to be invariant under rotation, and hence it cannot depend on $\dot{\bm{Q}}$ but rather on $\overset{\circ}{\bm{Q}}$ instead. They are both related as
$$
\overset{\circ}{Q}_{ij} = \partial_t Q_{ij} + v_k\partial_k Q_{ij} - \Omega_{ik}Q_{kj}+Q_{ik}\Omega_{kj},
$$
where $\Omega_{ij} = \frac{1}{2}(\partial_i v_j-\partial_j v_i)$. Alternatively \cite{re:chaikin95} one may argue that there needs to be a reversible coupling between $\dot{\bm{Q}}$ and the velocity $\bm{v}$. This leads to replacing $\dot{\bm{Q}}$ by $\overset{\circ}{\bm{Q}}$ in the reversible part of the evolution of $\bm{Q}$, Eq. \eqref{eq:qevol} \cite{re:hess75,re:brand94,re:kopf13}. 

\section{Tissue growth. Elasto nematic coupling}

Material evolution is due to elastic distortion, but also to tissue growth \cite{re:taber20}. One can write the total inverse deformation $\bm{W}$ as an elastic contribution $\bm{W}^{el}$ and a growth tensor $\bm{G}$ which is associated with spontaneous deformation at zero stress. The latter is due to both shape change induced by a change in the orientation distribution of actin fibers, and material creation through cell proliferation. We write $\bm{W}^{el} = \bm {G} \bm{W}$. This relation follows from the standard $\bm{F} = \bm{F}^{el} \bm{G}$ \cite{re:taber20}. This separation provides for the direct coupling through orientation and growth in $\bm{G}$ and elastic strains in $\bm{W}^{el}$. We also consider an incompressible cellular medium. Mass is not conserved due to cell proliferation, but the cellular medium is approximately incompressible (of constant density) \cite{re:rodriguez94}. Under these conditions, ${\rm det}(\bm{W}^{el}) = 1$, which implies that $\det(\bm W) =1/\det(\bm G)$. 
Since $\bm{L} = \dot{\bm{F}} \bm{F}^{-1}$, conservation of mass implies ${\rm Tr}(\bm{L}) = {\rm Tr}(\bm{F}^{el}\dot{\bm{G}}\bm{G}^{-1} \bm{F}^{el,-1})$.

A possible model for growth involves a linearization around homeostatic equilibrium of the form \cite{re:taber20},
$$
\dot{\bm{G}} = \gamma (\| \bm{T} \| - T_{0}) \bm{G},
$$
where $\gamma > 0$ is a constant growth rate, and $T_{0}$ the magnitude of the homeostatic stress. Further, in order to incorporate shape change under reorientation (due to remodeling) we adopt the following form of the growth tensor \cite{re:kopf13} $\bm{G} = \bm{I} + \alpha \ \delta \bm{Q}$, where $\alpha$ is a coupling coefficient (that can have either sign), and $\delta {\bm Q} = \bm{Q} - \bm{Q}^{(0)}$, the difference relative to some reference fiber orientation. Note that the nematic tensor order parameter $\bm{Q}$ is traceless. With this choice, ${\rm Tr}(\bm{L}) = d \gamma (\| \bm{T} \| - T_{0})$ in $d$ spatial dimensions.

The free energy is now a combination of elastic and nematic free energy $f = f_{el} + f_{n}$. We assume simply linear elasticity for the tissue
\begin{equation}
f_{el} = \frac{1}{2 \rho} \bm{E}^{el}: \mathcal{C} : \bm{E}^{el},
\label{eq:linelast}
\end{equation}
where $\mathcal{C}$ is the tensor or elastic moduli, and $\bm{E}^{el}$ is the Lagrangian strain tensor $\bm{E}^{el} = \frac{1}{2}(\bm{F}^{el,T}\bm{F}^{el} - \bm{I})$. Note that the choice of $\mathcal{C}$ depends on the symmetry of the reference state. In the experiments that we use in Sec. \ref{sec:discussion} to estimate parameters, the reference state is uniaxial, and hence the tensor $\mathcal{C}$ is that of a uniaxial or transverse isotropic phase \cite{re:brand94,re:chaikin95}. However, and in order to better illustrate the nature of the coupling terms in the theory, the small deformation analysis of Sec. \ref{sec:small_deformation} assumes an isotropic reference state instead. Nevertheless, the elastic energy in Eq. \eqref{eq:linelast} is written in a fully geometrically nonlinear form, and it includes a coupling to the orientational order parameter: The Lagrangian strain is
\begin{equation}
\bm{E}^{el} = \frac{1}{2} \left( (\bm{F}\bm{G}^{-1})^{T} (\bm{F} \bm{G}^{-1}) - \bm{I} \right)  = 
\bm{E} - \alpha (\delta \bm{Q} \bm{E} + \bm{E}\delta\bm{Q}+\delta \bm{Q}) + \mathcal{O}(\alpha^{2})
\label{eq:lstrain}
\end{equation}
which explicitly depends on the nematic tensor order parameter.

\section{Isotropic medium and small deformation}
\label{sec:small_deformation}

We consider in this section the limit of small deformation $\| \bm{U} \| \ll 1$ and small $\alpha$. This means that $\bm F \approx \bm I+\bm U$ such that 
\begin{eqnarray*}
    2\bm E = \bm F^T\bm F-\bm I = \bm U +\bm U^T \equiv 2\bm{\varepsilon},
\end{eqnarray*}
where we denote $\bm{\varepsilon}$ as the linear strain tensor. 

In order to illustrate the effects of the coupling terms, we also consider an isotropic and incompressible medium, so that the elastic energy is
\begin{equation}
f_{el} = \mu \bm{\varepsilon}^{el}: \bm{\varepsilon}^{el,T}
\end{equation}
where $\mu$ is its shear modulus. 
In this linear regime, Equation \eqref{eq:lstrain} reduces to 
\begin{equation}\label{eq:lstrain_linear}
\bm{\varepsilon}^{el} =
\bm{\varepsilon} - \alpha (\delta \bm{Q} \bm{\varepsilon} + \bm{\varepsilon}\delta\bm{Q}+\delta \bm{Q}),
\end{equation}
where the elastic strain is $\bm{\varepsilon}^{el}$. Thus, 
\begin{equation}\label{eq: fel_small}
    f_{el} = \mu \left[ \bm{\varepsilon} \bm{\varepsilon}^{T} - \alpha \left(\bm{\varepsilon} \delta \bm{Q} + \delta \bm{Q} \bm{\varepsilon}^{T} + \bm{\varepsilon} \bm{\varepsilon}^{T} \delta \bm{Q} + \delta \bm{Q}\bm{\varepsilon}\bm{\varepsilon}^{T}   + 2 \bm{\varepsilon} \delta \bm{Q} \bm{\varepsilon}^{T} \right)\right].
\end{equation}

Given Eq. \eqref{eq:ericksen}, the Ericksen stress for small deformation reduces to,
\begin{eqnarray}
T_{ij}^{R} & = & \frac{\partial f_{el}}{\partial \varepsilon_{ij}} + \rho f \delta_{ij} -\frac{\partial (\rho f)}{\partial(\partial_{j} Q_{mn})} \partial_{i}Q_{mn} \nonumber \\
\bm{T}^{R} & = & 2 \mu \bm{\varepsilon} - 2 \mu \alpha \delta \bm{Q} - 4 \mu \alpha (\bm{\varepsilon} \delta \bm{Q} + \delta \bm{Q} \bm{\varepsilon}) + \rho f \bm{I} + K \nabla  \bm{Q} \otimes \nabla \bm{Q},
\label{eq:ericksen_sd}
\end{eqnarray}
using the free energy given in Eq. \eqref{eq:ldg}. The last term in the right hand side also has the contraction over the tensor ${\bm Q}$. We note that there are two types of corrections to the stress. The second term on the right hand side is the same as for a nematic elastomer \cite{re:warner07}. A local change in nematic order induces distortion, which in turn modifies the elastic response (\lq\lq soft elasticity \rq\rq) (the coefficient $\alpha$ can be positive or negative depending on the material). The third term leads to a similar contribution, but it is formally of quadratic order. The next term is a consequence of tissue compressibility in the form of isotropic growth. The last term is the well known Ericksen stress in liquid crystals, which arises directly from nematic elasticity. Contributions at ${\mathcal O}(\alpha^{2})$ in Eq. \eqref{eq:lstrain} neglected here correspond to uniaxial elasticity.

Under normal stretching assays or physiological conditions, the tissue is expected to be in quasistatic elastic equilibrium given its current state of fiber orientational order (remodeling by actin polymerization takes place on a scale of seconds), and the volume occupied. We therefore write our first, coupled, governing equation as
\begin{equation}
\partial_{j} T_{ij}^{R} = 0, \quad {\rm or,} \quad \partial_{j} \dot{T}^{R}_{ij} =0.
\label{eq:eleq}
\end{equation}

We next turn to the equation of motion for the $\bm Q$ tensor using Eq. \eqref{eq:qevol}. 
Given $f_{el}$ from Eq.~\eqref{eq: fel_small} and the free energy from Eq. \eqref{eq:ldg}, one has for the right hand side of Eq. \eqref{eq:qevol}
\begin{equation}
\frac{\delta \mathcal{F}}{\delta \bm{Q}} = - \alpha \mu \left( 2 \bm{\varepsilon} + 4 \bm{\varepsilon} \bm{\varepsilon}^{T} \right) - K \nabla^{2} \bm{Q} - a \bm{Q} - b(\bm{Q}\bm{Q}) + c {\rm Tr}(\bm{Q}^{2}) \bm{Q}.
\label{eq:molfield}
\end{equation}

Thus, the evolution equation for $\bm Q$ coupled with mechanical equilibrium in two dimensions can be written as
\begin{eqnarray}\label{eq:small_deformation_eqs}
    \overset{\circ}{Q}_{ij} &=& \alpha\mu \Gamma (2\varepsilon_{ij}+4\varepsilon_{ik}\varepsilon_{jk})+\Gamma\left[K\nabla^2 + a - c{\rm Tr}(\bm{Q}^{2})\right]Q_{ij} + \Gamma b Q_{ik}Q_{kj}\\
    \partial_j T_{ij}^R &=& 0\\
    T_{ij}^R &=& 2 \mu \varepsilon_{ij} - 2 \mu \alpha \delta Q_{ij} - 4 \mu \alpha (\varepsilon_{ik} \delta Q_{kj} + \delta Q_{ik}\varepsilon_{kj})  + K (\partial_i  Q_{kl}) (\partial_j Q_{kl}).
\end{eqnarray}

\subsection{Steady-state small perturbation}
To examine more closely the feedback between nematic order and elasticity, we solve for the steady state 
resulting from a small perturbation $\|\delta \bm Q\|\ll 1$ relative to the equilibrium nematic order $\bm Q^{(0)}$ with $\|\bm Q^{(0)}\| = Q_0$ at zero stress, which the magnitude $Q_{0} = \sqrt{a/c}$ (we assume here and below that $b=0$). 
Thus, the reference state with uniform nematic $Q_{ij}^{(0)}$ corresponds to $T^{R0}_{ij} = \varepsilon_{ij}^{(0)} =0$. 
We also neglect cell proliferation as a source of deformation (${\rm Tr}(\bm{L}) = 0)$. In a time scale of seconds to minutes, the medium can be taken as effectively incompressible. 

Let us consider a uniformly perturbed state $\delta \bm Q, \ \delta \bm T^R$ due to an applied (uniform) strain $\delta\bm\varepsilon$. The constitutive relation between these perturbations  is
\begin{eqnarray*}
     \delta T_{ij}^R &=& 2\mu (\delta\varepsilon_{ij} - \alpha \delta Q_{ij}).
\end{eqnarray*}
To find the steady state equation for $\bm{Q}$ tensor equation, we need to linearise the cubic term,  
\begin{eqnarray*}
    {\rm Tr}(\bm{Q}^{2}) Q_{ij} &=& Q_{kl}Q_{kl}Q_{ij} =\\
    (Q_{kl}^{(0)}+\delta Q_{kl})(Q_{kl}^{(0)}+\delta Q_{kl})Q_{ij} &=& (Q_0^2+\delta Q_{kl}Q_{kl}^{(0)} +Q_{kl}^{(0)}\delta Q_{kl} +\mathcal O(\delta Q^2))(Q_{ij}^{(0)}+\delta Q_{ij})\\
    &=& (\delta Q_{kl}Q_{kl}^{(0)} +Q_{kl}^{(0)}\delta Q_{kl})Q_{ij}^{(0)}+Q_0^2\delta Q_{ij}+\mathcal O(\delta Q^2)
\end{eqnarray*}
Thus, the equation for $\delta\bm Q$ follows as
\begin{eqnarray}
     c(\delta Q_{kl}Q_{kl}^{(0)} +Q_{kl}^{(0)}\delta Q_{kl})Q_{ij}^{(0)}= 2\alpha \mu\delta\varepsilon_{ij}, 
\end{eqnarray}
using that $Q_0^2 =S_0^2 = a/c$.
Notice that the perturbation depends both on the magnitude and the orientation of the reference nematic field relative to the applied strain. We write the equation above for the two independent components $\delta Q_{xx}$ and $\delta Q_{xy}$. First, we evaluate 
\begin{eqnarray*}
\delta Q_{kl}Q_{kl}^{(0)} &=& \delta Q_{xx}Q_{xx}^{(0)} +\delta Q_{xy}Q_{xy}^{(0)}+\delta Q_{yx}Q_{yx}^{(0)}+\delta Q_{yy}Q_{yy}^{(0)}\\
&=&2(\delta Q_{xx}Q_{xx}^{(0)}+\delta Q_{xy}Q_{xy}^{(0)})
\end{eqnarray*}
Thus, 
\begin{eqnarray*}
\delta Q_{kl}Q_{kl}^{(0)}+Q_{kl}^{(0)}\delta Q_{kl} &=& 4(Q_{xx}^{(0)}\delta Q_{xx}+Q_{xy}^{(0)}\delta Q_{xy})
\end{eqnarray*}
Hence, the two coupled equations read as 
\begin{eqnarray*}
     4c(Q_{xx}^{(0)}\delta Q_{xx}+Q_{xy}^{(0)}\delta Q_{xy})Q_{xx}^{(0)} &=& 2\alpha \mu\delta\varepsilon_{xx}\\
     4c(Q_{xx}^{(0)}\delta Q_{xx}+Q_{xy}^{(0)}\delta Q_{xy})Q_{xy}^{(0)} &=& 2\alpha \mu\delta\varepsilon_{xy}
\end{eqnarray*}
or equivalently, by arranging terms, 
\begin{eqnarray*}
     (Q_{xx}^{(0)})^2\delta Q_{xx}+ Q_{xx}^{(0)}Q_{xy}^{(0)}\delta Q_{xy} &=& \frac{\alpha \mu}{2c}\delta\varepsilon_{xx}\\
     Q_{xx}^{(0)}Q_{xy}^{(0)}\delta Q_{xx}+(Q_{xy}^{(0)})^2\delta Q_{xy} &=& \frac{\alpha \mu}{2c}\delta\varepsilon_{xy}
\end{eqnarray*}

For a system in three dimensions, the nematic tensor order parameter can be written as \cite{re:selinger16} $Q_{ij} = S (\frac{3}{2} n_{i}n_{j} - \frac{1}{2} \delta_{ij})$, where the unit vector $\hat{\mathbf{n}}$ is the uniaxial director, and $S$ is the degree of uniaxial order. This definition is consistent with the definition of the structure tensor in Ref. \cite{re:gasser06}, except that the $\mathbf{Q}$ tensor is made traceless. An orientationally disordered system has $S = 0$, whereas the perfectly oriented case has $S = 1$. In order to compare our results with existing data on thin films of patterned microtissue (Sec. \ref{sec:discussion}), the analogous definition of the tensor order parameter in two dimenions is $Q_{ij} = S(2\hat n_i\hat n_j-\delta_{ij})$. In this case, the director can be written as $\hat {\bm{n}} = [\cos(\theta); \sin(\theta)]$, and $\bm{Q}$ only has two independent components given by 
\[Q_{xx} = S\cos(2\theta), \qquad Q_{xy} = S\sin(2\theta)\]
Thus, 
\begin{eqnarray*}
      \cos^2(2\theta_0)\delta Q_{xx}+ \sin(2\theta_0)\cos(2\theta_0) \delta Q_{xy} &=& \frac{\alpha \mu}{2c S_0^2}\delta\varepsilon_{xx}\\
     \sin(2\theta_0)\cos(2\theta_0)\delta Q_{xx} +\sin^2(2\theta_0)\delta Q_{xy}&=& \frac{\alpha \mu}{2c S_0^2}\delta\varepsilon_{xy}.
\end{eqnarray*}

\begin{figure}[t]
\center
\includegraphics[width=0.5\linewidth,valign=c]{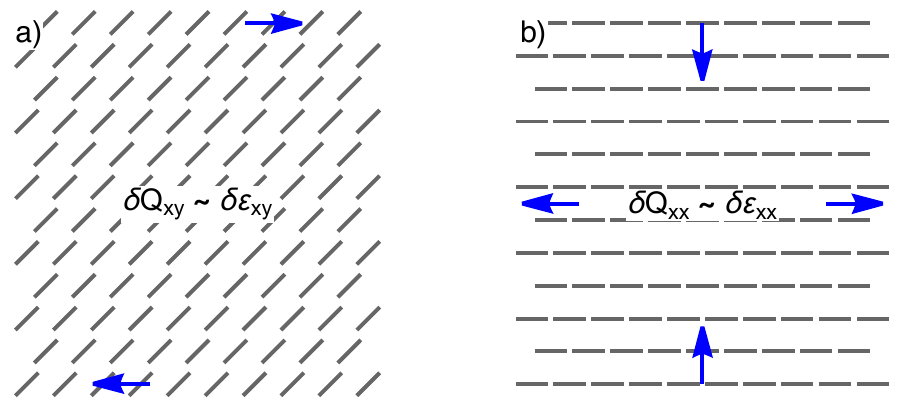}
\caption{Steady state orientation of the nematic director for a) simple shear from Eq.~\ref{eq:Qpert_xy} and for b) uniaxial strain from Eq.~\ref{eq:Qpert}.}
\label{fi:op_theory}
\end{figure}

For an arbitrary orientation $\theta_0$ of the reference state, we have that 
\begin{eqnarray*}
\left(\begin{matrix}
\cos^2(2\theta_0) & \sin(2\theta_0)\cos(2\theta_0)\\
\sin(2\theta_0)\cos(2\theta_0) &  \sin^2(2\theta_0)  
\end{matrix}\right) \left(\begin{matrix}
\delta Q_{xx}\\
\delta Q_{xy}
\end{matrix}\right) = \frac{\alpha\mu}{2c S_0^2}\left(\begin{matrix}
\delta\varepsilon_{xx}\\
\delta\varepsilon_{xy}
\end{matrix}\right)
\end{eqnarray*}
Notice that the matrix determinant vanishes for any $\theta_0$, 
meaning the system is degenerate and reduces to only one equation 
\begin{equation}
\cos(2\theta_0)\delta Q_{xx}+ \sin(2\theta_0) \delta Q_{xy} = \frac{\alpha \mu}{2c S_0^2}\left[\cos(2\theta_0)\delta\varepsilon_{xx}+\sin(2\theta_0)\delta\varepsilon_{xy}\right].
\end{equation}
There are two reference nematic orientations $\theta_0$ for which these perturbations decouple: 
\paragraph{a) $\theta_0= 0$}which corresponds to the reference nematic state aligned with the direction of (compression/extension) chosen to be along $x$ as shown in Fig.~\ref{fi:op_theory} a). In this case, 
\begin{equation}
      \delta Q_{xx} = \frac{\alpha \mu}{2c S_0^2}\delta\varepsilon_{xx}
\label{eq:Qpert}, \qquad \delta T_{xx}^R = 2\mu \left(1 - \frac{\alpha^2}{2c S_0^2}\right)\delta\varepsilon_{xx}.
\end{equation}

\paragraph{b) $\theta_0= \pm\pi/4$} corresponds with the reference orientation at $\pm\pi/4$ degrees aligned with the shearing direction along $x$ as shown in Fig.~\ref{fi:op_theory} b). Then,  
\begin{equation}\label{eq:Qpert_xy}
\delta Q_{xy} = \frac{\alpha \mu}{2c S_0^2} \delta \varepsilon_{xy}, \quad 
     \delta T_{xy}^R = 2\mu \left(1 - \frac{\alpha^2}{2c S_0^2}\right)\delta\varepsilon_{xy}.
\end{equation}

\section{Discussion and conclusions}
\label{sec:discussion}

The central results of our analysis, described in Sec. \ref{sec:small_deformation}, are the contributions to the reversible stress from terms that depend the on the nematic tensor order parameter, Eq. \eqref{eq:ericksen_sd}. Terms that are linear in $\delta \mathbf{Q}$ are formally equivalent to those describing soft elasticity in conventional nematic elastomers, whereas those that are bilinear in $\delta \mathbf{Q}$ and $\bm{\varepsilon}$ are of higher order in the small deformation setting. Such a response results from elemental shape changes under reorientation when the element of volume is under no stres, and it is proportional to the coupling coefficient $\alpha$ which we estimate below. 

We consider the results of experiments designed to probe the elastic response of micro patterned tissue (Madin–Darby canine kidney tissue, or MDCK) in the recently developed cellular micro-biaxial stretching system (C$\mu$BS) \cite{re:cook23}. Rectangular, thin tissues were fabricated with a variety of aspect ratios, and stretched under controlled conditions. Internal stresses were determined by in situ traction force microscopy, and fiber orientation distributions measured by confocal microscopy in tissue stained for F-actin. The results corresponding to two particular aspect ratios are examined here: A square (1:1) micro tissue which is seen to be largely un-oriented, and responds as an elastically isotropic medium, and a rectangular (2:1) tissue which displays significant nematic order along the long direction, order that changes under the stretching protocols \cite{re:cook23}.

\begin{figure}[t]
\centerline{
\includegraphics[width=0.35\linewidth,valign=c]{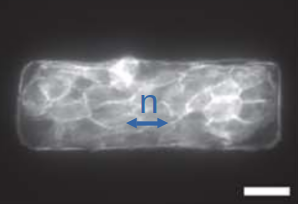}\hspace{0.5cm}
\includegraphics[width=0.6\linewidth,valign=c]{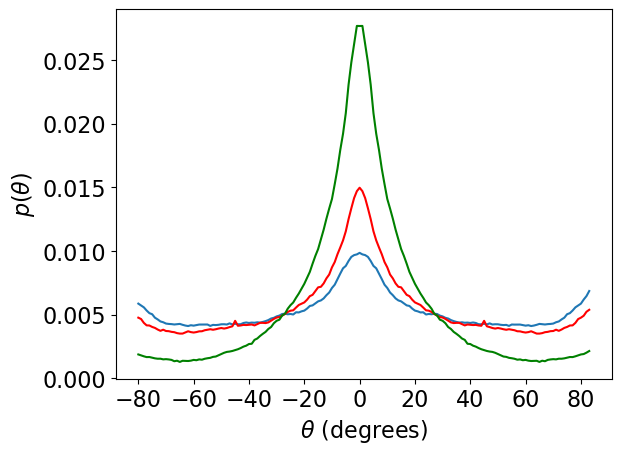}}
\caption{Left: Bright field image of MDCK tissue patterned in a rectangular shape (2:1 aspect ratio) showing its F-actin network. The predominant orientation along the $x$ axis (left to right) is denoted by the director $\hat{\mathbf{n}}$. Right: Distribution of fiber orientation $p(\theta)$, with orientation relative to the $x$ axis. Shown are the distributions for aspect ratio 1:1 (blue), aspect ratio 2:1 (red), and aspect ratio 2:1 under a 23\% uniaxial strain along the $x$ direction (green). The respective order parameters are $S = 0.41, 0.50$ and 0.75.}
\label{fi:op}
\end{figure}

Figure \ref{fi:op} shows the measured orientation distribution following the protocol of Ref. \cite{re:cook23}. Shown are two control distributions for aspect ratios 1:1 (blue) and 2:1 (red), and then the distribution when the tissue has been stretched by 23\% along the $x$ direction (parallel to the director). Orientation increases significantly. The uniaxial order parameter can be computed as $S = \langle \frac{3}{2} \cos(\theta)^{2} - \frac{1}{2}\rangle$, where the average is taken over the measured distribution $p(\theta)$. We find $S = 0.41$ for 1:1 aspect ratio, $S = 0.50$ for 2:1 aspect ratio but no strain, and, $S = 0.75$ for 2:1 under uniaxial strain. Therefore $\delta Q_{xx} = \delta S = 0.25.$

The strain reported in the experiments is $\delta \varepsilon_{xx} = 0.23$. The resulting stress change $\delta T_{xx}^{R}$ is more difficult to obtain as the experimental results (Fig. 9 of \cite{re:cook23}) consider long duration runs of 24 hours, while tissue stretching per se only took 15s. During the remainder of the experiment cell proliferation dominates the deformation response. We therefore estimate the stress from the discontinuities at zero time shown in Fig. 9C (AR1 or 1:1 aspect ratio) and Fig. 9E (AR2, or 2:1 aspect ratio) in \cite{re:cook23}. First, for that 1:1 sample, we estimate $\delta T_{xx} \approx 0.20 kPa$ which yields an estimate of the shear modulus $\mu \approx 0.54 kPa$ for for the assumed isotropic medium. This is within range of known Young modulus of MDCK \cite{re:bruckner17}. In the oriented case, we estimate $\delta T_{xx}^{R} \approx 0.17 kPa$. From Eqs. \eqref{eq:Qpert}, we obtain $\alpha \approx 0.16$.


The estimates just given have assumed uniform distributions of fiber orientational order and strain to match existing mechanical results. However, both are gross oversimplifications as neither one is uniform in the micro tissue. Nonuniformities in fiber orientation will result in nonuniform stresses which need to be spatially resolved in order to quantitatively describe tissue biomechanics. In the case of the C$\mu$BS experiments of Ref. \cite{re:cook23}, an analysis of traction force microscopy results reveals highly nonuniform distributions of stresses both in the isotropic AR1 and oriented AR2 tissues. Figure 4, for example, shows that the local values of tissue stress range from almost zero to 350~kPa. These local variations are of the same order as the tissue average values reported, which have been used in the estimates above.

Since we have confined the analysis of Sec. \ref{sec:small_deformation} to corrections at ${\mathcal O} (\alpha)$, uniaxial elasticity at ${\mathcal O}(\alpha^{2})$ plays no role. However, this is bound to be an oversimplification as, clearly, the longitudinal stiffness of actin fibers must contribute to the elastic response of an oriented micro tissue (see, e.g., the analysis in Ref. \cite{re:win17} for a single cell). The nonlinear analysis of Sec. \ref{sec:transport_model} contains those terms as well. However, there is no experimental information at present that would allow us to separately analyze nematic elastomer like response from uniaxial elasticity. Additional experiments along these lines are currently ongoing.

\section*{Acknowledgments}
This collaboration is funded by the Norwegian Centennial Chair Program, a program jointly funded by the Government of Norway and the University of Minnesota Foundation. The research of PA is supported by the National Science Foundation, contract CMMI 2230435, and JV by DMR 2223707.

\bibliographystyle{unsrt}
\bibliography{Qet}

\end{document}